# Trail systems self-organize through individual route optimization

Brendan Wallace[1] (corresponding author) and Andrew Berdahl[1,2]
Email: b.wallace@hey.com

[1]Center for Quantitative Sciences, University of Washington
[2]School of Aquatic and Fishery Sciences, University of Washington

# Abstract

As animals, including humans, walk, they compact the vegetation and ground beneath them, reducing the cost of movement for subsequent travelers. From this positive feedback emerge networks of trails which allow for efficient movement and store generational-scale information about the environment. Models simulating this process show how individuals following local cues produce trail systems with many of the features of real-world networks. However, little has been done to investigate walkers with more complete knowledge of their environment and the abilities to plan their behavior intelligently. Here we introduce a simple general model for trail formation, based on minimizing cost of movement by individuals. As in real environments, our simulations produce a striking variety of complex trail networks. We show that these trail systems optimally connect space according to the spatial needs of the walkers, that the geometry of those trail systems matches theoretical predictions and that the systems they generate may fall into multiple attracting states. As technological advances allow us to quantify trail systems across diverse habitats and species, our model provides a theoretical foundation for empirical research on animal trails and for understanding the emergence of human transportation networks, with implications for urban design.

# Introduction

Stepping off the sidewalk to take a shortcut through the grass leaves an indent, making it easier for the next person to walk the same way. Over time, repeated use feeds back on itself, forming a more established trail. In urban design and popular culture this kind of trail is called a "desire path," and thought to indicate a mismatch between the built environment and people's needs[1–3]. In the 1960s, Michigan State University took advantage of this process when they expanded their campus without immediately adding new sidewalks. The university let students walk through the grass to form their own footpaths, only later returning to pave the trail system that emerged[4] (Fig 1).

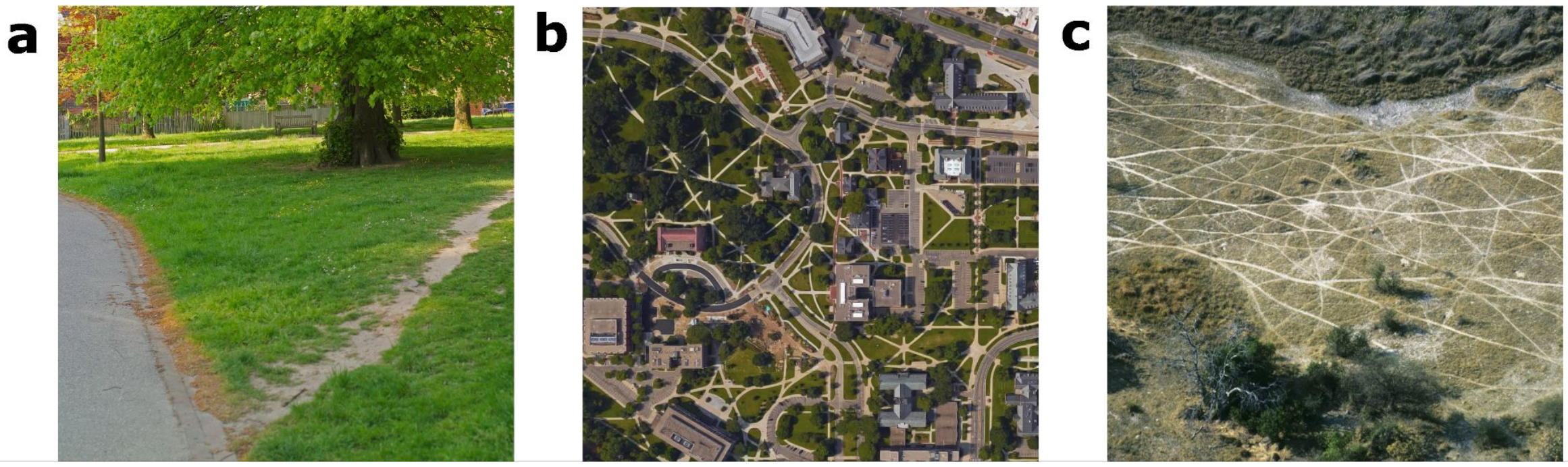

**Figure 1. Desire paths and trail systems.** (a) Stepping off a sidewalk to cut through the grass leaves an indentation, making it easier to walk the same way. Over time, positive feedback can form a trail; sometimes called a desire path or a game trail**2** (photo: Alamy). These trails can form entire networks. (b) At Michigan State University, walking paths were allowed to form organically across the lawn before being paved**4** (photo: Maxar Technologis, USGS). (c) Similar networks emerge from animal movements, as illustrated by game trails in the Okavango Delta, Botswana (photo © Tony Heald / naturepl.com).

Such self-organized trail systems, or networks, are ubiquitous across a wide range of ecological and sociological domains[5]. Across diverse terrestrial habitats, animals establish trail systems that connect important locations within their home ranges[6–8] and facilitate repeated movements along seasonal migration routes[9]. Grazing livestock tracks form terracettes on hilly landscapes, and facilitate energy-efficient movement even across flat terrain[10–12]. Camera trap data reveals that a single trail can be used, and thus maintained, by a wide range of species[13]. At smaller spatial scales, eusocial insects form networks of chemical trails through a similar positive-feedback process, laying pheromone as they travel while preferentially following the trails laid by others[14–18].

Animal trail systems represent a distributed offline collective information storage system that spans individuals, species and generations[19,20]. A trail system summarizes countless movement decisions made by many individuals, potentially over long periods of time. In a sense, such trails are an environmentally-mediated analogue of the "many wrong" principle – rather than averaging movement preferences simultaneously, repeated use integrates movement decisions serially over many individuals and generations, reinforcing successful routes while unsuccessful routes fade away[21–23]. Instead of encoding the entire landscape, a trail network stores a compressed representation of where movement is repeatedly advantageous. Access to this compressed information may reduce the cognitive demands of navigating complex environments. Because this information resides in the environment rather than within individual brains, it can accumulate over timescales far exceeding the lifetime of any individual, allowing movement decisions to be shared across generations and even species[24].

Viewed through this informational lens, we can see that trail systems have ecological implications far beyond simply making movement more efficient. A naive individual entering the landscape can immediately exploit information, analogous to a public good, using existing trails to locate resources and travel efficiently between them. Such information may be especially important during recolonization after local extirpation. Further, by concentrating a large portion of movement onto an approximately linear subset of the broader two-dimensional landscape, trail systems fundamentally alter interaction rates among mates, predators, prey and pathogens[7,25,26].

Theoretical work to understand trail systems falls under the broad umbrella of stigmergy — collective behavior mediated through modification of the environment — and has tended to focus on eusocial insects[14,27–29]. As such, modeling efforts have focused on reinforced random walks wherein agents respond to local gradients to preferentially move over places that other agents have been, with limited knowledge of their larger environment[15–17]. Broadening the perspective to include humans, Helbing et al. introduced an agent-based model in which individuals move through a two-dimensional space, increasing the attractiveness of the ground at their current position, and updating the direction of their movement based on the location of their target as well as the gradient of ground-attractiveness at their position[30–32]. Other theoretical models have compared trail formation mathematically to water flow[33], electric flow[34], and slime mold[35], or explored the shape and structure of optimal transportation networks in general[36–38]. These models often demonstrate that simple, local behavioral rules and logic are sufficient to create trail systems that resemble those observed empirically.

However, gaps remain in our theoretical understanding of trail formation. Existing models largely treat trails as attractive features that bias movement through local interactions with the environment. Yet many animals, including humans, possess detailed spatial knowledge of their surroundings, remember distant locations, and plan directed routes toward known goals following least-effort paths[11,12,19,39]. Under these circumstances movement may be better viewed as the outcome of global route planning than local attraction alone. This distinction suggests an alternate formulation. Rather than moving up gradients of attractiveness, walkers may instead move along routes that minimize the cumulative cost of reaching their destination, where cost represents any combination of energetic expenditure, discomfort, social constraints, or obstacles.

Here we introduce a model based on minimization of a generic travel cost and explore scenarios in which agents operate with full knowledge of their environment (Figure 2). In our model, walkers navigate a two-dimensional space, planning routes that minimize the cumulative travel cost to their destinations. Local travel costs decrease as walkers pass and return towards a background level when paths are not used. As walkers traverse the landscape, a system of paths quickly emerges and may continue to evolve on longer time scales. Collectively, the walkers form efficient transportation networks that minimize travel costs for a given amount of trail, with some conditions giving rise to multiple distinct long-term network configurations. The geometry of these networks is sensitive to the relative cost of travelling on and off trails.

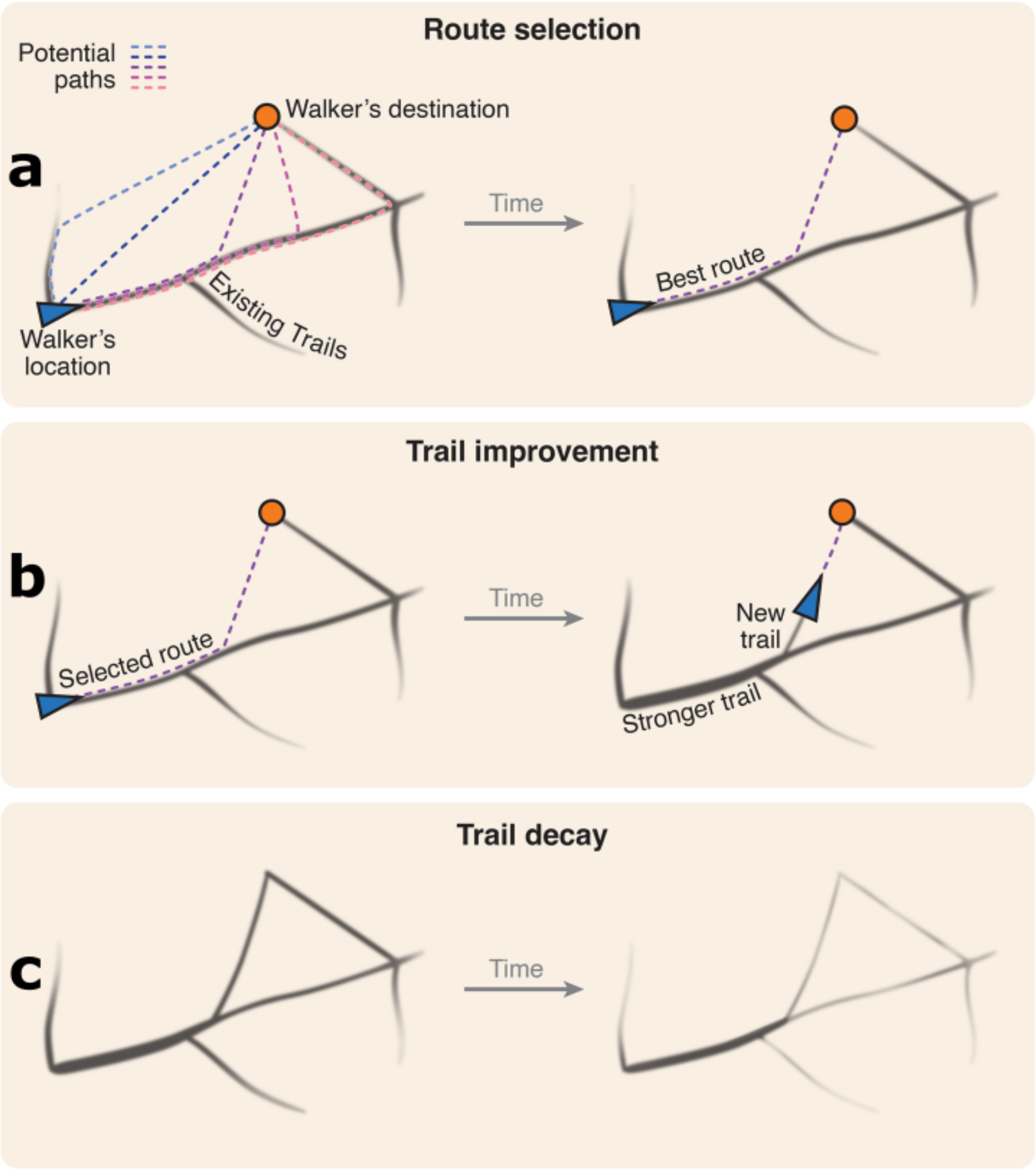

**Figure 2. Model of trail formation.** We model self-organized trail network formation using an agent-based model. (a) Walkers compute the least-cost path to their destination. (b) As walkers move, they decrease the cost of ground at rate $\tau_+$, constructing a network of low-cost trails. (c) Unused trail slowly disappears at rate $\tau_-$, returning the ground to its original high-cost state.

## Results

**Qualitative trail system patterns depend on configurations of destinations**

Rather than observing a generic trail system across simulated scenarios, from each scenario emerges a distinct pattern of trails (Fig 3). When destination points are fixed, the trail system connects all the points in a Steiner-tree-like manner, with trails merging and splitting (Fig 3a). When walkers alternate between a fixed central point and dynamic random points along the exteriors, the trails form a branching tree shape (Fig 3b). When walkers move between dynamic, randomly placed locations along the left and right edge of the arena, mimicking long-range, direction-based, migration, an array of disjoint linear trails slowly coalesce into a tree-like shape connecting the walls (Fig 3c). In our uniformly Random Movement scenarios, after a walker arrives at a point, the point is removed and placed in a new uniformly random position, so that the trail system tries to connect the entire space. When boundaries are fixed, the trail system forms a central ring or cross with branching tendrils (Fig 3d, Fig 4a). When boundaries are periodic, the trails may fill the space evenly or converge in a (potentially offset) central location (Fig 3e, Fig 4b).

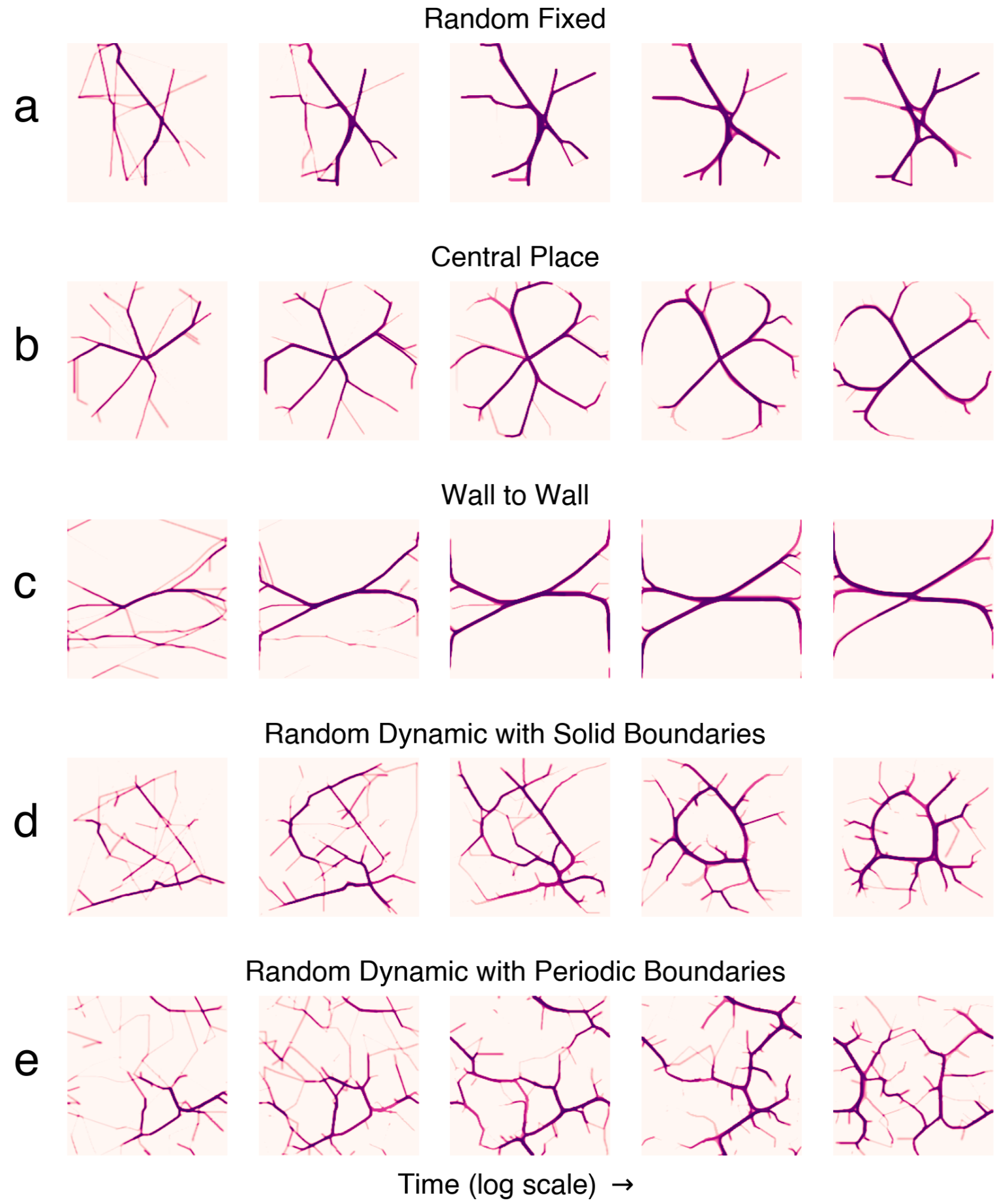

**Figure 3. Walkers form intricate trail systems.** We explore a range of scenarios – logic of how points are arranged and where walkers visit – to see what kind of trail system the walkers produce. a) Fixed points: walkers move between fixed points of interest, creating a network of trails connecting those points. b) Central-place forager: walkers alternate between random, shifting points on the edge of a circle (foraging locations) and the center of the circle (home), creating a tree-like structure. c) Transit: walkers transit the space, moving between random, shifting points on the left and right walls. d) & e) Random movement: walkers move between shifting points placed randomly throughout the entire space, constructing a network that spans the space. Boundaries at the edges of the simulation are solid (individuals cannot pass through) for all scenarios except that shown in e), in which all boundaries are periodic and c), in which the top and bottom boundary are periodic. Snapshots are taken at doubling periods (400, 800, 1,600, 3,200 and 6,400 time steps) to show how structure develops over time, with $\tau_+ = 0.04$, $\tau_- = 0.001$, $L = 200$ and $C_{ground} = 4$ for each simulation.

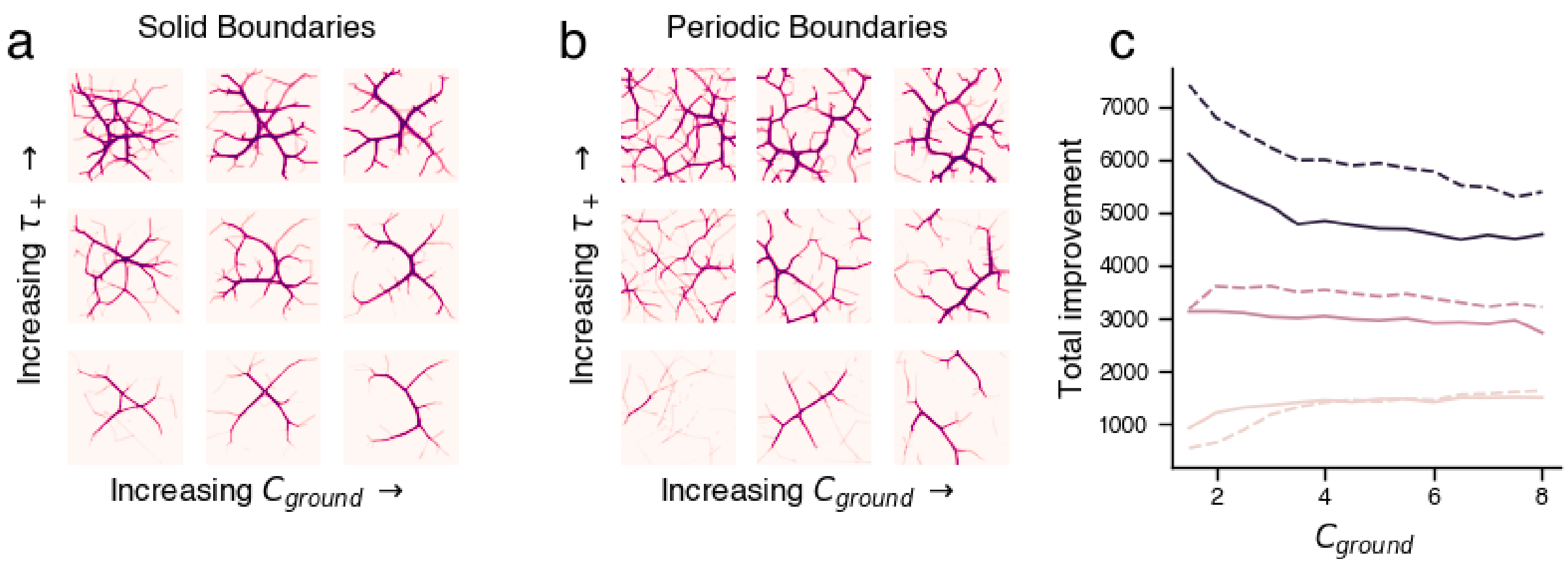

**Figure 4. Trail systems respond to model parameters.** We show trail systems after 10,000 time steps for random shifting points with solid (a) and periodic (b) boundary conditions, and varying trail budgets and trail preferences. As we increase $\tau_+$, walkers create more expansive systems. As we increase $C_{ground}$, walkers generally create more constrained networks with higher-degree angles, but at low trail budgets (bottom row), this is partially reversed and walkers create more expansive networks. In (c), we run simulations with corresponding parameters and averaged over 4 trials to display these trends quantitatively. Solid lines indicate solid boundaries, dashed lines indicate periodic boundaries. Top rows and black lines use $\tau_+ = 0.16$; middle rows and purple lines use $\tau_+ = 0.08$; bottom rows and peach lines use $\tau_+ = 0.04$.

**Environmental properties determine expansiveness of trail system**

Walkers construct more extensive trail systems when trail is created easily (high $\tau+$) and decays slowly (low $\tau-$); and less extensive trails systems when trail is created less easily (low $\tau+$) and decays rapidly (high $\tau-$) (Figure 4a & b, top-to-bottom). This is to be expected: each timestep, each of $W$ walkers creates $\tau+$ new trail, and $\mathcal{L}\tau_-$ recovers, where $\mathcal{L}$ is the amount of existing trail. Setting these equal and solving for $\mathcal{L}$,

$$\mathcal{L}^* = \frac{W\tau_+}{\tau_-} \quad \text{(equation 1)}$$

is the maximal amount of trail walkers can sustain – we call this quantity the *trail budget*.

When the relative cost of traveling on untouched ground to traveling on established trail is high (Fig 4a & b right columns), the model reflects animals moving through high-cost environments – i.e., dense vegetation, or forming paths through snow; when this ratio is low (Fig 4 a & b left columns), the model reflects people walking across a park green or animals moving through a grassy area. In general, we find that as we increase this cost ratio, walkers form less extensive trail systems, and trail segments meet at angles that are closer to orthogonal. This is because walkers will more preferentially use existing trails that lead less directly towards their targets, and they will stay on these trails for longer before departing at close to a 90-degree angle. However, when the trail budget is particularly low, some of these effects are reversed. Trail generally does not form at all in these cases. But as the cost ratio increases, walkers are incentivized to detour further to reuse nascent trail, leading to a higher degree of positive feedback which allows walkers to construct *more* extensive systems (Fig 4 a&b bottom rows, left-to-right; Fig 4c).

**Walkers allocate trail budget efficiently**

We next ask how *efficient* the trail systems are. Here we only consider scenarios with fixed destinations, and consider the trail system as a transportation network. To quantify its efficiency, we consider the dual-optimization 'transportation problem,' of simultaneously minimizing the travel cost between a set of points, and the total amount of infrastructure (*i.e.* amount of trail) constructed[6,38]. We run a series of otherwise-identical simulations in which we systematically vary the trail budget by changing $\tau_+$. After a fixed amount of time, we measure both the total amount of trail and the pairwise cost of travel between all points.We find that, when measured this way, trail systems fall neatly along a curve in this solution space that neatly resembles a pareto-front. Infrastructure costs cannot be decreased by moving to a different solution point without increasing transportation costs, and vice versa. Collectively, the population efficiently adapts to the presented trail budget by making a tradeoff between length of trail and overall travel costs (Fig 5).

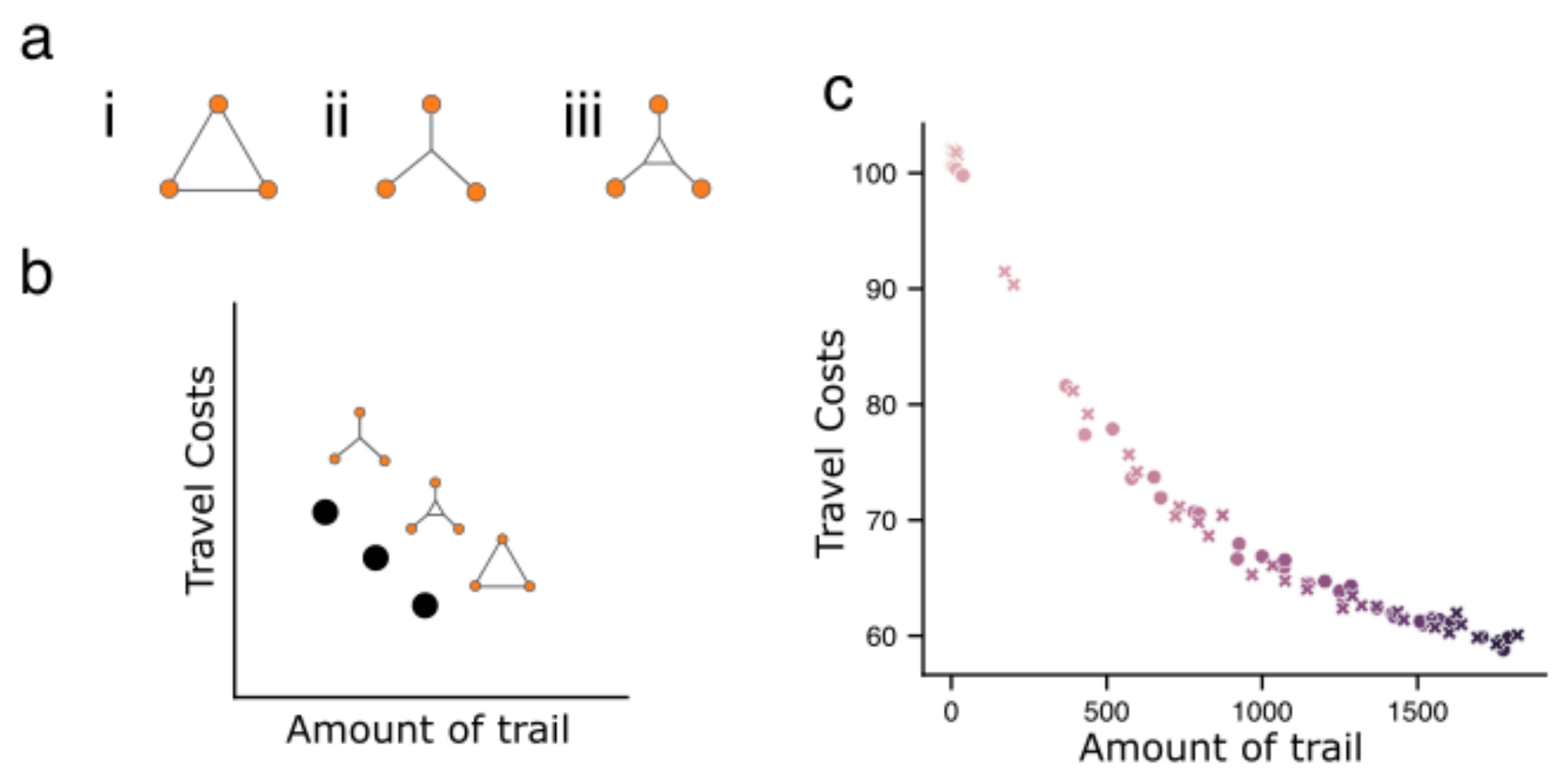

**Figure 5. Optimality of network solutions.** We quantify the efficiency of transportation networks in reference to the "transportation problem": the dual-optimization task of connecting a set of points in space with trail to minimize travel costs between them while simultaneously minimizing the amount of infrastructure (trail) used[6]. (a) A solution may use maximal (i), minimal (ii), or an intermediate amount (iii) of trail to connect the same three points. (b) When plotted in space with the amount of trail used along one axis, and travel costs (average lowest cost of travel between points*)* along the other, different solutions fall along a curve. (c) We run 128 simulations with $L = 100$ for 10,000 time steps, with one of two identical configurations of 32 points (indicated by x or o) of interest, fixing $\tau_{-} = 0.001$ while varying $\tau_{+}$ from 0.01 to 0.2 (darkening color indicates increasing $\tau_{+}$) to explore different constraints on the amount of trail. We find that solutions trace a Pareto-efficient curve, where no solution could improve upon the travel costs of any other without requiring more trail, and vice versa.

**Network geometry varies predictably with strength of trail preference**

We next investigate the relationship between walkers' preference for staying on the trail and the topography of the system they form. When the cost ratio is high, walkers tend to produce more minimal systems whose trails meet at angles closer to 90 degrees. This is consistent with intuition and commonly observed patterns. Walking through a field after a fresh snow, one tends to follow existing footprints as much as possible; conversely, walking through a grassy field one may follow a trail a way, but leave it as soon as the straight path towards one's destination diverges significantly.

To quantify this relationship we can find that the angle at which one should leave a straight trail to walk directly towards one's target, is

$$\arccos(\frac{C_{trail}}{C_{ground}}),$$ (equation 2)

where $C_{trail}$ is the cost of walking on the trail and $C_{ground}$ is the cost of walking on the untouched ground. In our simulations, as in the real world, trail segments are not always straight or homogeneous. Still, this *critical angle* is fixed for a given ratio of costs, and we predict a relationship between observed angles in the trail system and this theoretical critical angle: trails meeting and turning below this angle should be common, greater angles should be rare.

To verify, we run a series of otherwise-identical simulations, varying the ratio of ground-to-trail costs (Fig 6a). After 5,000 time steps, we measure the internal geometries of the trail systems produced (Fig 6b). Specifically, for each set of points, we first compute the *least-cost* path to move between the points, and then we measure the difference in headings along this path minus the heading directly towards the destination point at the end of the path. We aggregate these heading differences and characterize their distribution by finding the 90th, 95th, and 99th percentiles (Fig 6c). The tail of these angle distributions is closely related to the predicted *critical angle* (eq 2) for each scenario. We find a strongly linear relationship in each case (Fig 6d).

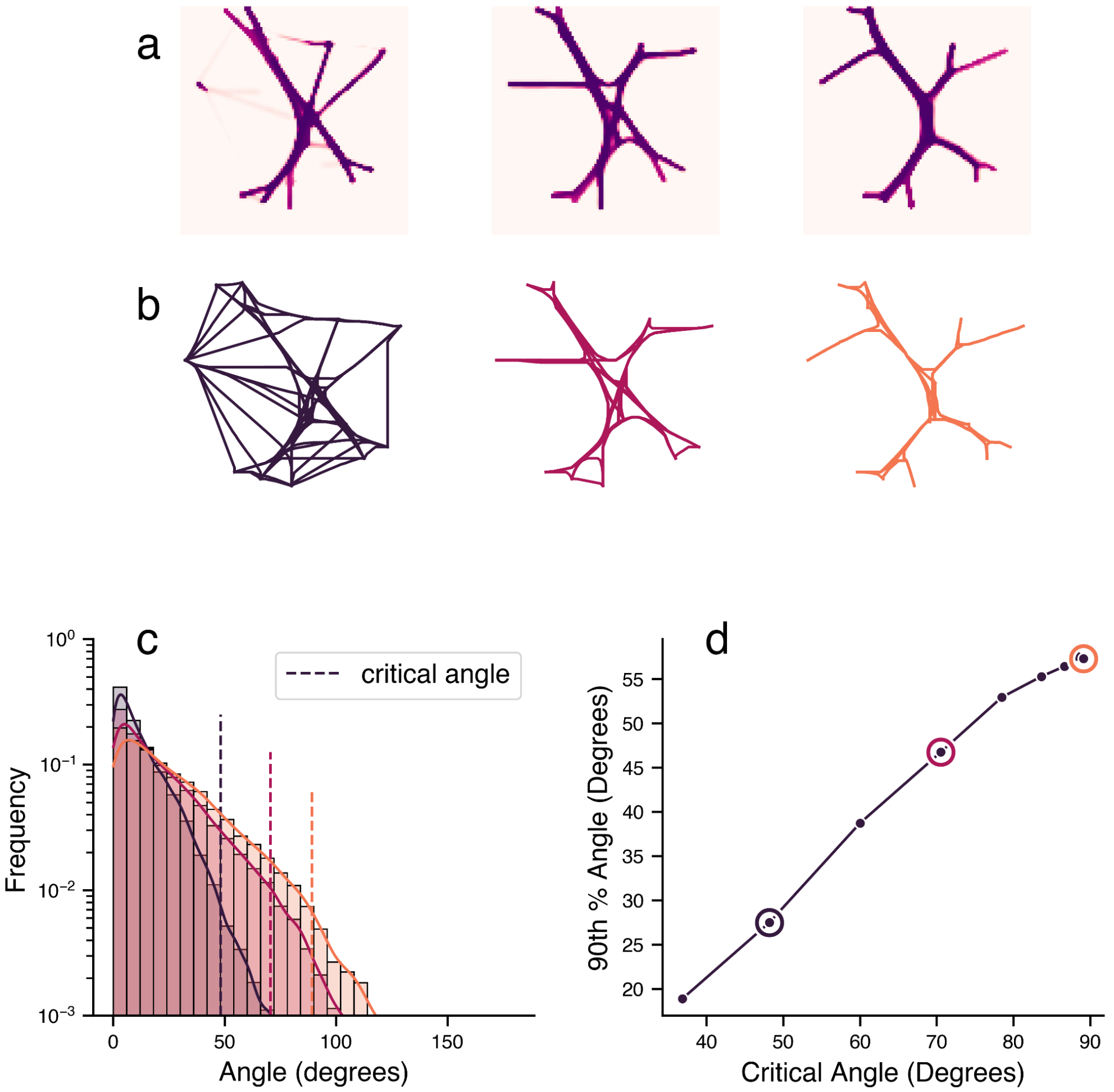

**Figure 6. Relationship between ground costs and network angles.** We investigate the relationship between trail preference (cost of moving across untrampled ground vs cost of moving on trail) and geometry of the network produced. (a) We run simulations across a range of trail preference values (highlighted examples left to right with $C_{ground}$ = 1.5, 3.0, and 65.0) for 5,000 steps. (b) We compute all pairwise shortest paths between points of interest in each network. (c) Along each shortest path, we define the "angle of deviation" as the difference in angle between the current trajectory on the path and the vector towards the target at the end of the path. We aggregate angles along all shortest paths to produce a distribution of angle-of-deviation for each value of trail preference. (d) We plot the 90th percentiles from these distributions against the theoretical critical angle $\arccos(C_{ground}/C_{trail})$, revealing a strongly-linear relationship. $\tau_{+} = 0.1$, $\tau_{-} = 0.001$, and $L = 100$ for all simulations in this figure, with $C_{ground}$ ranging from 1.25 to 65.0, with 4 trials at each of 6 different location configurations (2 configurations with 10 POIs, 2 with 15, and 2 with 20).

**Systems may exhibit multiple attracting states**

Finally, if the simulation parameters and point configuration are the same, will the same network form each time? In other words, is this process deterministic, or will one of multiple potential trail systems emerge to connect a given set of points? To investigate, we fix all model parameters and run 128 replicas of a scenario, randomizing the initial locations of the walkers to introduce randomness. We let each simulation run for a fixed length of time, producing 128 final networks (Fig 7ai). We then perform a dimensionality reduction (see Methods) on the set of final systems to visualize the dynamics of the systems forming over time in a smaller dimensional space, allowing us to identify how many distinct trail systems exist (Fig 7ai&ii). We perform this analysis across a range of model parameters and point configurations. For some point configurations, we observe multiple distinct and stable systems and early differences can tip the process into one of several dynamic attracting basins (Fig 7b). For other point configurations, all observed trail systems develop towards a single shape (Fig 7c), though it is possible there are additional unobserved trail solutions.

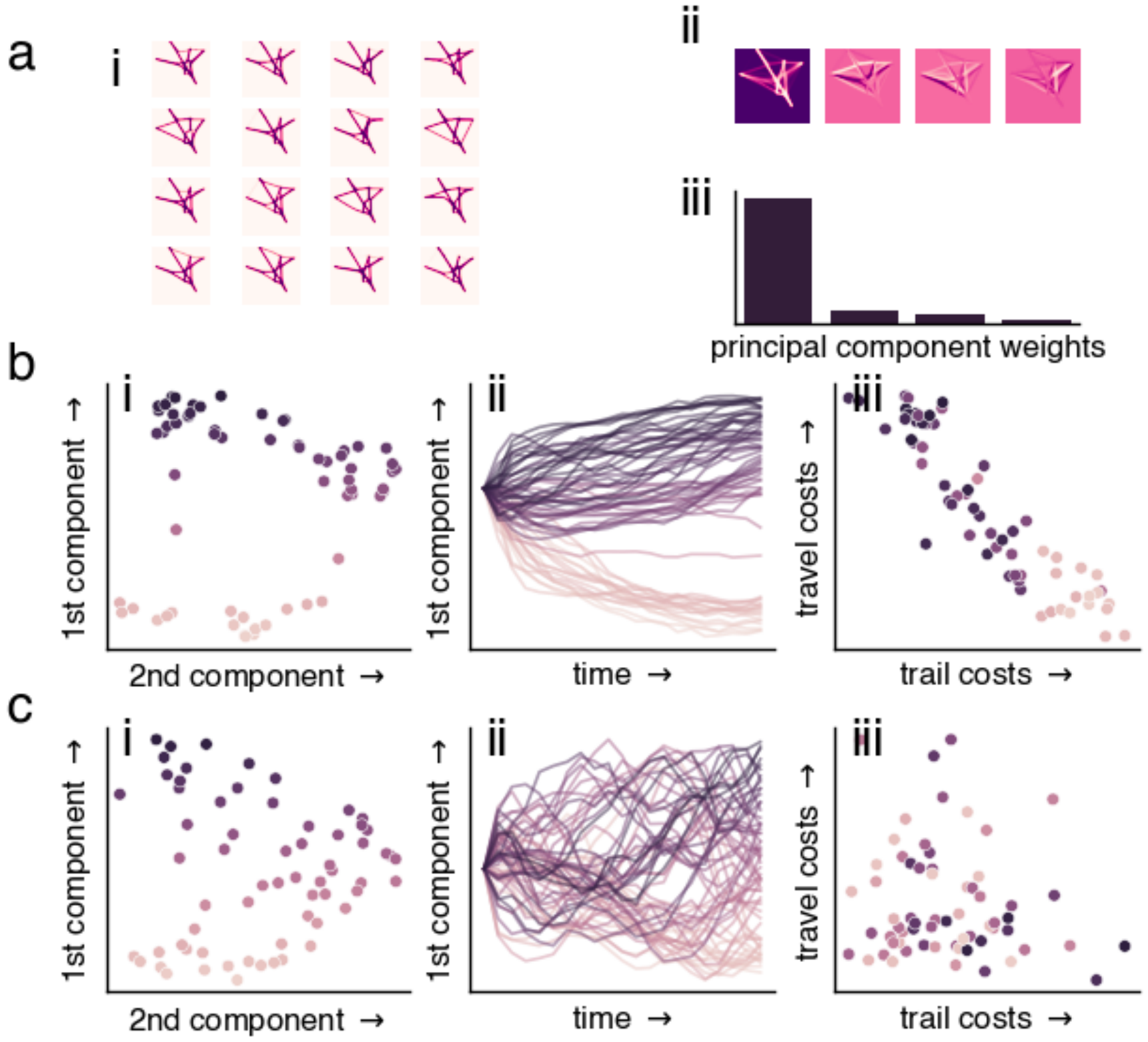

**Figure 7. Dynamics and stability of network solutions.** (a) We investigate the stability and dynamics of the trail-network system by running 64 simulations with identical parameters while introducing noise through different starting positions of the walkers. We construct a matrix of the final trail systems of each replica, and perform singular-value decompositions. The matrix decomposes into interpretable "eigen-networks" and their respective weights, giving an indication of how many unique solutions exist in the system. (bi) For some sets of parameters, final network systems decomposed into their first two principal value components cluster into multiple distinct solutions. (bii) When we plot the evolution of these components over time, we observe multiple stable-appearing basins of attraction. (biii) Finally, when we measure the efficiency of the final networks (see Fig 5), we see that clustering in decomposition-space corresponds to functionally different solutions. (ci-iii) For other sets of parameters, network systems don't cluster in decomposed space. In (i-iii), points and lines are colored by their final weights of their first component in the decomposition in (a), further highlighting the stability in (b) and lack of structure in (c). Simulations in (b) use $\tau_- = 0.0005$, $\tau_+ = 0.005$, 8 locations, and $C_{ground} = 2.0$. Simulations in (c) use $\tau_- = 0.001$, $\tau_+ = 0.01$, 8 locations, and $C_{ground} = 8.0$. All simulations use $L = 100$, and 20,000 time steps.

## Discussion

Our model differs from other models of trail formation by inverting the concept of *attractiveness,* to one of *cost.* In previous models, walkers are drawn towards areas of high comfort (or pheromones, etc.) as they move towards a target. In our model, walkers move towards their target while minimizing the accrued *cost*. This reformulation has multiple advantages. *Cost* in our model captures many concepts well, such as energetics, discomfort, social stigma, obstacles. Following local gradients is computationally simple but can be subject to undesirable edge-case behavior, as when attractiveness in different directions cancel with one another and a walker ignores both; whereas animals as simple as flies make concrete directional decisions with their movement.[40] Moreover, while both *cost* and *attractiveness* can be quantified, the cost *along a path* can be minimized meaningfully and used algorithmically, while the *attractiveness* along a path cannot be used the same way. This is because a path attempting to *maximize* attractiveness would spend infinite time looping over attractive parts of the trail and never reach its destination (and this is a corollary of the fact that cost-minimizing algorithms fail when negative costs are present in a graph).

Constructing a trail network is fundamentally an optimization problem. The mathematical history of this problem dates back to the early 1800s with the Euclidean Steiner tree problem, which asks how to connect a set of points in a plane with the shortest-length network. Solutions of this problem are essentially minimal spanning trees, called Steiner trees, and it is known to be NP-hard[6,41].

Our model allows us to ask how well decentralized walkers, interacting only through environmental modification, collectively solve this NP-hard optimization problem. Previous work demonstrates that emergent trail systems have consistent allometric scaling properties[42], and minimize trail length in a "pro-steiner" direction, that is, when compared to a fully connected *travel-minimizing* system, emergent trail systems minimize trail in the direction of a Steiner tree[6,30–32]. We confirm this general finding in our model, and go further by demonstrating that active walkers construct a full *range* of pro-steiner deviations to fill out a pareto-front of solutions (Fig 5). No one walker is concerned with the optimality of the larger system, yet from this collective of individuals, all acting in their immediate best interest (i.e., taking cost-minimizing paths), a fine-grained and efficient system emerges.

Our findings have implications for making inferences about animal behavior from observed trail systems. For example, multiple studies have made a similar observation that animals' trails appear less densely in open areas than in heavily wooded areas, and then drawn similar conclusions that animals are *frequenting* open areas less than wooded areas[43,44]. Our findings suggest an alternative explanation: that landscapes with high cost of movement – e.g. heavily wooded area – are simply more conducive to the development of emergent trails than those with relatively low cost, due to the higher degree of positive reinforcement motivating animals to reuse emerging trails rather than forge new ones. As another example, our results indicate a direct relationship between the internal geometries of trail systems and the cost energetic difference – energetic or otherwise – between walking along trails and walking along ground. This relationship is present in the trail-branching angles of leaf-cutting ants[45], and in general, if we can measure the angles present in trail systems created by animals, we can make direct quantitative hypotheses about their energetics, leading to predictions about their movement and behavior that may be useful in conservation and other applications.

Our results also have implications for understanding human transportation systems, from footpaths through parks to cities' entire road networks. Helbing observed that pedestrians consistently form new shortcuts (desire paths) across the grass when an existing paved path is at least 30% longer than the potential shortcut[46]. This 30% detour rule is equivalent to a ground-to-path cost ratio of 1.3. Using this value in (equation 2) we find a *critical angle* of about 40°, leading to the prediction that desire paths should tend to leave sidewalks at about 40° angles, and that where paved paths run at angles of *greater* than 40° relative to a destination, desire paths are likely to form.

We opened with a story of sidewalks at Michigan State University. After the university let students form trails through the grass, were they correct to pave them? Our results suggest the trail system they formed did efficiently connect the buildings on campus,reflecting the travel needs of the students. However, it may not have been a unique solution and perhaps the system formed wasn't necessarily final, as trails continuously evolve. Similarly, had the grass been taller or wetter, students may have formed a more minimal trail system with longer walking detours, and the opposite if the grass was shorter or more easily trampled.

Zooming out further, the dynamics at play in our model may have shaped some of our large-scale transportation infrastructure. Paths formed by elephants have been incorporated into road networks.[47] Even more directly, the *traditional model of development* holds that businesses and foot traffic along a road inspire further investment in the same place: improved road infrastructure, more businesses and increased foot traffic; creating a positive feedback loop.[48,49] This suggests that the logic in our model should apply to the road network of cities. We think these systems deserve further study, as their structure and dynamics provide a lens into a simplified version of tradeoffs relevant to transportation planning[50], and may hold insights into the structure of existing cities and how cities emergently develop.

We have developed an intentionally minimalistic model and as such it involves assumptions and limitations that are important to discuss. Our model assumes agents to have perfect knowledge of the landscape. In reality, animals traverse a landscape with varying knowledge of that landscape as well as varying cognitive abilities to make optimal decisions and imperfect estimates of the energetic costs they will pay. A natural extension of our model could involve more limited landscape knowledge and decision-making abilities. Our model assumes a homogenous and fixed (other than trail formation) cost environment. Future work could also incorporate heterogeneous (in space and/or time) costs, landscape topography[12], as well as physical obstacles in the environment. We model a single, homogeneously behaving, species, when in real systems trails are used by multiple species, which may have different travel costs and levels of landscape modification as they pass by. We do not include resources explicitly in our model. Future versions that also incorporate a dynamic resource field could link costs and benefits of movement in a novel and insightful way. Our model incorporates a computationally-expensive search algorithm which limits the feasibility of running the model on larger grids. Limiting the search algorithm so that walkers move optimally only within a limited vision horizon, and rely on a heuristic approach beyond that, would decrease the computational costs of the model and allow for investigation of situations in which animals act with limited information.

With rapid advances in satellite- and UAV-based remote sensing[51], together with computational methods for extracting features such as trail systems from imagery[52,53], we expect quantitative data on empirical trail systems to be widely available in the near future[12,54–56]. Simultaneously, the interest in human and animal movement is growing rapidly. The widespread availability of cellphone GPS data has driven an explosion of research on human mobility and general patterns of movement[57]. Likewise, advances in GPS tracking collars[58,59], overhead filming and automated tracking[44,60,61], and open-source initiatives such as Movebank[62] have dramatically increased the availability of animal movement data. As quantitative data on trail systems become increasingly available, so too will the value of predictive theories of how they emerge and how animals move within them. We hope our model provides a useful step toward this goal.

## Methods

**Simulation Model Details**

In our model, agents follow the least-cost paths between target points of interest across a 2-dimensional space, improving (decreasing the cost of) regions of the space as they move. Improvements slowly decay so that the ground reverts back to its original high-cost state.

### The Space

While the agents can move in any direction, the ground is quantized into a finite (typically either 100 x 100 or 200 x 200) number of 1x1 unit-distance patches, and ground improvement (i.e. creation of trail) is tracked individually in each square.

### Points of Interest

Points of interest (POIs) are added to the simulation and targeted by the walkers differently depending on the model scenario.
**Random Fixed**: We choose N (typically 8 or 16) random points anywhere in the space initially, which remain fixed for the duration of the simulation. Walkers are assigned target POI by random, and reassigned randomly upon reaching a target.
**Central Place:** One POI is placed at the center of the simulation. Walkers begin in the center. For each walker, we generate a target POI at a fixed radius from the center in a random position (i.e. a random position on a circle). When a walker reaches its target outer POI, that POI is deleted and the walker is reassigned to target the central POI. Walkers alternate in this way between the central POI and the randomly generated peripheral POIs.
**Transit:** Walkers start in random positions. Each walker is assigned a randomly-generated target POI on the left wall of the simulation. When a walker reaches a target on one wall, that POI is removed and the walker is assigned a new randomly-generated POI on the opposite wall.
**All Random:** Walkers start in random positions. Each walker is assigned a randomly-generated POI anywhere in the simulation. When a walker arrives at a POI, that POI is removed and the walker is assigned to target a new randomly-generated POI. Boundaries at the edges of the simulation are solid (individuals cannot pass through) for all scenarios except for All Random with Periodic Boundaries.

## Least-cost Paths

After a walker is assigned a target POI, it computes the least-cost path to arrive at that POI. Then it follows that path until reaching its destination, even if changes to the landscape occur during transit. Upon reaching its destination, the walker is assigned a new POI, computes a new least-cost path, and begins moving during the next time step. Paths are not restricted to discrete directions; walkers can move in any direction as they traverse the space. The cost of a path is the sum of the lengths of the path over each patch times the cost (per unit distance) of traversing that patch.

## Search algorithm

Our model computes least-cost paths with a variation of an algorithm for approximate shortest-path computation over a polyhedron by *Kanai and Suzuki*.[63] The algorithm is as follows:

- Mark the vertices of the polyhedron as nodes in a graph. (Here, the polyhedron is the $L \times L$ grid of patches, and the corners of the grid patches the vertices.).
- Introduce edges to connect vertices across faces of the polyhedron.
- Until an end criterion is met:
  - Introduce additional nodes uniformly at intermediate positions between vertices along original edges of polyhedral faces. In our model, we add 3 additional nodes in-between existing nodes.
  - Introduce additional edges connecting the new nodes.
  - Compute the least-cost path from a start vertex to an end vertex over the graph using Dijkstra's algorithm[64].
  - Prune all nodes and edges not touched by the least-cost path from the graph.
  - (End criterion: the algorithm ends when the cost of the least-cost path found decreases by less than 0.1, or after completing three iterations.)
- Return the final least-cost path.

Our model's path routing requires approximating a solution to the *weighted region shortest path* problem, which is known to be technically unsolvable[65], and very computationally expensive even to approximate to a high degree of optimality[66]. We use this approximation algorithm, which is reported to be 100-1000 times faster than comparable versions, and within 0.4% of optimal[63,66].

## Time and Walker Speed

Time is discretized for the sake of implementing the simulation on a computer. Each simulation update (time step) corresponds to 1 unit of time, which we define to be how long it takes for a walker to move the length of one patch of ground, which is one unit of length.

## Ground Improvement and Decay

Each patch of ground stores a level of improvement between 0 (no improvement) and 1 (maximal improvement). This is translated into cost linearly, such that

$$cost = C_{trail}(\text{improvement}) + C_{ground}(1 - \text{improvement})$$.

Walkers increase improvement as they move, at the rate of trail improvement $\tau_+$. Each timestep, the distance *d* a walker traversed over a patch (at most 1 per time step) triggers a total of $\tau_+$ times *d* improvement centered in this patch, with gaussian blur with standard deviation of 1 into the surrounding 48 patches (the 7 x 7 grid of patches), where distance is computed between the centers of the patches. When boundaries are solid, any improvement that would be applied outside the boundaries of the simulation is discarded. When boundaries are periodic, improvement is applied across the boundaries to the other side(s) of the simulation. Ground improvement is capped at 1.

Improvement decreases at trail recovery rate $\tau_-$. After walkers move and ground is improved, all ground patches' improvement values decrease by $\tau_-$, to a minimum of 0.

## Other Details

We use 10 walkers for all simulation results presented here. We run most simulations for 10,000 time steps.

## Model parameters

| Symbol | Description | Value / Equation |
|---|---|---|
| $\tau_+$ | Trail improvement rate | [0.01, 0.2] |
| $\tau_-$ | Trail decay rate | {0.001, 0.0005} |
| $C_{trail}$ | Trail traversal cost | 1.0 |
| $C_{ground}$ | Ground traversal cost | [1.0, 65.0] |
| | Trail preference | $C_{ground}/C_{trail}$ |
| $W$ | Number of walkers | 10 |
| $\sigma_{\tau_+}$ | Spatial scale / width of trail improvement | 1.0 |
| $L$ | World side-length (in number of squares) | {100, 200} |

**Table 1. Model parameters.**
We show model parameters, descriptions and their value ranges explored in the paper.

Model parameters and their ranges are shown in Table 1.

### Analyses

## Optimality of Trail Solutions

We find the pairwise cost of travel between all points by explicitly computing all least-cost paths using the same search algorithm used by walkers in the simulation.

We perform the analysis multiple times, using 8, 16, and 32 randomly-generated fixed points (Random Fixed scenario). For each number of points, we consider two different configurations, as well as two different max costs: 2.0 and 8.0, for a total of 12 analyses. For each analysis, we keep trail recovery fixed and vary trail improvement rate $\tau_+$ from 0.01 to 0.2 by increments of 0.01, running two simulations at each value. We run each simulation for 10,000 steps, using 10 walkers on a 100x100 size grid.

## Network Geometry

We perform the analysis with 10, 15 and 20 POIs. We consider max costs of 1.25, 1.5, …, 65.0 ( $1.0 + 2^i$ with $i \in -2, \dots, 6$ .). We use $\tau_+ = 0.01$ and $\tau_- = 0.001$ , on a L=100 grid with 10 walkers. We consider two different configurations of POIs for each setting.

## Dynamics

We perform the analysis multiple times, using 8 and 16 randomly-generated fixed points (Random Fixed scenario), periodic and solid boundary conditions, and two different random configurations of points for each combination. We additionally consider a low ground cost (2.0) and high ground cost (8.0), and faster trail creation ( $\tau_+ = 0.01$ , $\tau_- = 0.001$ ) and slower trail creation ( $\tau_+ = 0.005$ , $\tau_- = 0.0005$ ). We use 10 walkers on a 100x100 grid.

For each analysis, we run 64 simulations with identical parameters and POI configurations, with different initial positions of walkers set randomly. We run each simulation for 20,000 steps. We treat each of 64 final grids of improvements as 100x100 raster images with values between 0 and 1. For the dimensionality reduction, we perform Principal Component Analysis by finding the singular value decomposition of these 64 rasters, first converting them into a 10,000 x 64 matrix[67]. Denoting the matrix of rasters as A, this gives the decomposition,

$$A = U \Sigma V^T.$$

To visualize how trail systems evolve in the column space of $V^T$ , we follow a similar procedure, taking snapshots of the trail systems every 20 time steps and treating them as raster images. We convert each 100x100 image to a 10,000x1 vector, and multiply by $V$ (the inverse of $V^T$ , because $V$ is unitary) to find the weights of each singular component over time.

## Funding

The authors declare no relevant funding.

## Data Availability

The code and data for this model are available at https://github.com/brendanwallace/trailsystems

## Author Contribution

B.W. and A.B. conceived the model. B.W. wrote simulation code and performed analyses. B.W. wrote the initial draft, B.W. and A.B. reviewed and edited.

## Competing Interests

The authors declare no competing interests.